\documentclass[twocolumn,prb]{revtex4}
\usepackage{graphicx}
\usepackage{subfigure}
\usepackage{bm}
\usepackage{amssymb}
\usepackage{wasysym}
\usepackage{color}

\newcommand{\Ket}[1]{\vert \, #1 \, \rangle}

\renewcommand{\phi}{\varphi}
\renewcommand{\epsilon}{\varepsilon}

\begin{document}

\title{Nuclear Spin Dynamics in Double Quantum Dots: Fixed Points, Transients and Intermittency}

\author{M. S. Rudner$^{1}$, F. H. L. Koppens$^{2,3}$, J. A. Folk$^{2,4}$, L. M. K. Vandersypen$^{2}$, L. S. Levitov$^{5}$}
\affiliation{
$^{1}$ Department of Physics, Harvard University, 17 Oxford St., Cambridge, MA 02138\\
$^{2}$ Kavli Institute of NanoScience, TU Delft, PO Box 5046, 2600 GA, Delft, The Netherlands\\
$^{3}$ ICFO-Institut de Ciencies Fotoniques, Mediterranean Technology Park, 08860 Castelldefels (Barcelona), Spain\\
$^{4}$ Department of Physics and Astronomy, University of British Columbia, Vancouver, British Columbia V6T 1Z4, Canada\\
$^{5}$ Department of Physics, Massachusetts Institute of Technology, 77 Massachusetts Ave, Cambridge, MA 02139
}

\begin{abstract}
Transport through spin-blockaded quantum dots provides a means for electrical control and detection of nuclear spin dynamics in the host material.
Although such experiments have become increasingly popular in recent years,
interpretation of their results in terms of the underlying nuclear spin dynamics remains challenging.
Here we examine nuclear polarization dynamics within a two-polarization model that supports a wide range of nonlinear phenomena.
We point out a fundamental process in which nuclear spin dynamics can be driven by electron shot noise; fast electric current fluctuations generate much slower nuclear polarization dynamics, which in turn affect electron dynamics via the Overhauser field. 
The resulting intermittent, extremely slow current fluctuations account for a variety of observed phenomena 
that were not previously understood. 
\end{abstract}

\maketitle


The opportunity to study spin coherence and many-body dynamics 
in a controllable solid-state setting has inspired a wide range of experiments in a variety of materials such as GaAs vertically grown and gate-defined structures\cite{Hanson07},
InAs nanowires\cite{Pfund}, and $^{13}$C-enriched carbon nanotubes\cite{Churchill}.
In particular, electron transport through spin-blockaded double quantum dots\cite{OnoSB} constitutes a purely electrical means of probing and manipulating the dynamics of nuclear spins.
Such experiments have revealed complex dynamical
phenomena, including bistability and hysteresis\cite{Pfund,Churchill,OnoTarucha, Baugh}, switching\cite{Koppens,Reilly,Churchill}, slow transient build-up of current\cite{Koppens}, and slow oscillations\cite{OnoTarucha, Austing}.  

Despite wide interest in these phenomena and their importance for quantum information processing, progress in understanding them has been slow. 
While there is little doubt that nuclear spins in the host material play a crucial role, the lack of a direct 
 probe of nuclear spin dynamics requires their behavior to be inferred from electronic transport measurements.
To meet this challenge, theoretical modeling must be used to complement analysis of relevent features in transport data.  

\begin{figure}
\includegraphics[width=3.4in]{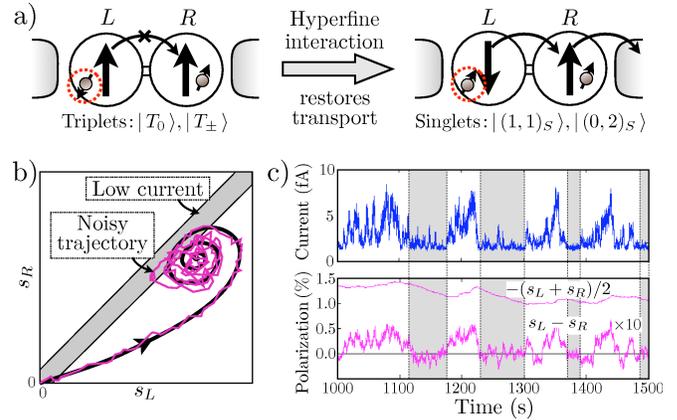}
 \caption[]{Nuclear dynamics and intermittent current fluctuations driven by shot noise of the spin-blockaded current through a double dot. 
a) Hyperfine spin-exchange with nuclei mediates transitions between two-electron triplet and singlet states, relieving blockade and producing dynamical nuclear polarizations (DNP) $s_L$ and $s_R$ in the left and right dots.
b) A phase portrait of DNP trajectories, with a fixed point (DNP steady-state) positioned near the main diagonal, where current is low due to spin-blockade (see Fig.\ref{fig:TT}). 
c) Typical simulated current trace showing the effects of steady state DNP fluctuations (pink traces in b and c), see appendix. 
Switching between quiet and noisy regions results from excursions into the dark stripe $s_L\approx s_R$, marked by the shaded regions in panel c). 
%
 }
\label{fig1}
\end{figure}

In previous work on spin dynamics in double quantum dots, simple models involving a single dynaical variable describing the total nuclear polarization have been used to explain the origin of feedback in this system\cite{RudnerDNP, Platero1}. 
Although such models can succesfully account for feedback-driven nonlinear phenomena such as bistability and hysteresis, 
the range of phenomena which they can describe is somewhat limited.
Here we expand the phase space of the model, and describe nuclear spin dynamics in terms of {\it two} dynamical variables, $s_{L}$ and $s_R$, corresponding to the independent nuclear polarizations in the left and right dots (see also e.g.~Ref.~\onlinecite{Platero}), thereby extending the range of phenomena that can be analyzed.
Time evolution is described by 
trajectories in a two-dimensional phase space $(s_L, s_R)$, which 
can exhibit complex dynamics 
including non-monotonic behavior, limit-cycles, or spirals, as illustrated in Fig.\ref{fig1} (also see Refs.~\onlinecite{DanonNazarov} and \onlinecite{Gullans} for additional examples of complex phenomena arising from two-polarization dynamics in other contexts).


Nuclear polarization dynamics in spin-blockaded dots is driven by carriers passing through the system\cite{OnoSB}. 
Each electron passing through the dot can produce a spin flip of the nuclei due to hyperfine exchange with nuclear spins in the host lattice, see Fig.\ref{fig1}a.
Dynamic nuclear polarization (DNP) arises when the up and down spin-flip rates are imbalanced, $\Gamma_+\ne\Gamma_-$ \cite{OnoTarucha, RudnerDNP}. 
Since the spin-flip rates in the two dots are in general different, their corresponding nuclear polarizations $s_L$ and $s_R$ have different time dependence, generating an asymmetry between the dots, $s_L\ne s_R$. 
As we shall see, such asymmetry is dramatically reflected in the time-dependence of electric current.


In this work we focus on the effects in nuclear polarization dynamics due to the shot-noise arising from the discreteness of 
carriers passing through the system. 
Electrons are injected into the system one-by-one, with random spin orientations.
While transiting through the dots, each such electron may exchange its spin with the nuclear subsystem. 
Crucially, these stochastic spin-flip processes comprise an {\it intrinsic} source of broadband noise that couples to nuclear dynamics. The intensity of this noise, which is proportional to the DC current, remains nonzero even when the average rates of up and down spin-flips are equal: $S\propto (\Gamma_++\Gamma_-)$. The resulting DNP fluctuations  
are relatively slow due to the large number of nuclear spins in the dots, $N\approx 10^6$, which requires many electrons to be transmitted through the system before DNP 
can change substantially.

Another important aspect of the double dot system is the complex relationship between the system's internal variables and measurable quantities, i.e.~between the nuclear polarization and electric current. Due to the resonant energy dependence of transition rates, current is sensitive to the alignment of energy levels via a number of external and internal variables (gate voltages, magnetic field, Overhauser fields in each dot, etc). 
Changes in the hyperfine spin flip rates feed back into DNP dynamics, giving rise to a variety of interesting nonlinear phenomena occurring on long time scales, exemplified in Fig.\ref{fig1}b,c.
Numerical simulations based on this microscopic model, which is described in detail below, demonstrate how the complex long timescale dynamics 
arise from the stochastic nature of electron transport.

\begin{figure}
\includegraphics[width=3.4in]{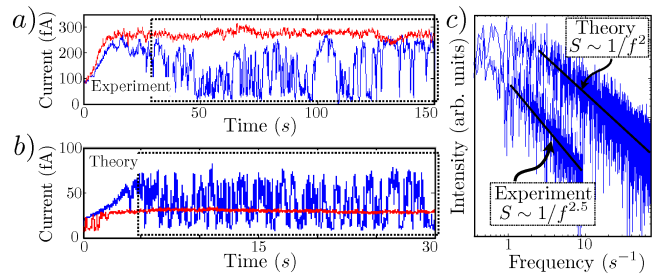}
 \caption[]{
Time-dependent current, from experiment \cite{Koppens} (a), and from simulation (b), 
showing a long build-up lasting several tens of seconds followed by steady-state intermittent fluctuations (blue traces).
Fluctuations can be suppressed by a small change of detuning (gate voltage) which moves a fixed point away from the sensitive region $s_L\approx s_R$ (red traces).
c)  Fourier spectra of the experimental and simulated noisy traces, evaluated in the marked steady-state regions.
The spectra display a roughly $1/f^\alpha$ dependence with $\alpha_{\rm exper.}\approx 2.5$ and $\alpha_{\rm simul.}\approx 2$. 
}
\label{fig:exp_traces}
\end{figure}

In particular, we find that the high frequency noise can drive intermittency in electric current resembling the multi-scale switching behavior observed in experiments, which will be discussed below. 
In dynamical systems, intermittency refers to the alternation of phases of apparently regular and chaotic dynamics\cite{Vassilicos}. 
Such behavior arises in many physical systems. For example, fluorescence intermittency, or blinking, is commonly observed in the optical repsonse of various nanoscale systems, such as large molecules or quantum dots, where it signals competition between the radiative and non-radiative relaxation pathways\cite{Stefani}. In our system, a commonly observed type of behavior is slow build-up followed by intermittent switching between ``quiet'' and highly fluctuating current states, illustrated in Fig.\ref{fig1}b,c and Fig.\ref{fig:exp_traces}a,b.

Throughout this paper, simulation results are compared to data from the measurements described in Ref.\onlinecite{Koppens}. 
Figure \ref{fig:exp_traces}a shows typical experimental current traces observed in the regime of moderate magnetic field ($B = 200\ $mT) and with a gate voltage setting where the electrostatic energy makes lowest two singlet states, one with one electron in each dot and the other with both electrons in the right dot, nearly degenerate (i.e.~near zero ``detuning'').  
In this case, these ``$(1,1)$'' and ``$(0,2)$'' singlet states are strongly hybridized by the tunnel coupling between the dots (see Fig.\ref{fig:TT}a).
The traces were taken after a long waiting
period which allowed the system to relax to equilibrium. Current
displays dynamics on a very long time scale, with a smooth transient
``slow build-up'' period lasting several tens of seconds followed by a
``steady-state'' featuring intermittent large amplitude fluctuations with a correlation time on the scale of seconds (blue trace). 
The fluctuations can be abruptly
suppressed by a relatively small change of detuning 
(red trace). Similar 
behavior was observed during slow sweeps of magnetic field (shown in Fig.\ref{fig:Sweeps}).
Similar-looking fluctuations were reported by Reilly et al.~as `blinking' of the Overhauser field measured in a double-dot which was repeatedly pulsed through a singlet-triplet level-crossing\cite{Reilly}. 
There, long timescale noise correlations were attributed to nuclear spin diffusion resulting from the dipole-dipole interaction. 
In contrast, below we describe a mechanism where diffusion of the net nuclear polarization is not driven by the conventional dipole-dipole mediated spin flips, but rather  
is driven by shot noise in the current passing through the system.

The rest of the paper is organized as follows.
In Section \ref{sec:TwoPol} we describe the physical mechanism of shot-noise-induced multiscale intermittent fluctuations of current.
Then in Section \ref{sec:Model} we present the mathematical description of our model for describing the time dependence of nuclear polarization and current in spin-blockaded double quantum dots. 
In Section \ref{sec:Results} we present the results of simulations based on the model described in Sec.\ref{sec:Model}, and compare with experimental data.
Finally, our conclusions are summarized in Section \ref{sec:Conclusions}.

\section{Transients and Intermittency in the Two-Polarization Model}
\label{sec:TwoPol}
%
%

A typical behavior, often seen in the data, is a relatively slow transient buildup of current after which the system enters an intermittent state, characterized by alternation of quiet and noisy behavior. 
Here we discuss the physics of how such behavior can arise naturally from the two-polarization model.
The key elements of the mechanism are summarized in schematic form in Fig.\ref{fig1}b, which shows a phase portrait of DNP in the $(s_L,s_R)$ plane. 
In our analysis we assume that, via the hyperfine interaction, dynamics are primarily controlled by two variables $s_L$ and $s_R$ that describe independent nuclear polarizations in the two dots.
The trajectories shown in Fig.\ref{fig1}b are  
obtained by applying the ideas of Refs.~\onlinecite{OnoTarucha} and \onlinecite{RudnerDNP} to the case of two coupled polarizations while ignoring noise; the fixed points associated with these trajectories describe steady-state DNP. 
%
%
Due to blockade of the triplet state $\Ket{T_0}$, current is low in the gray stripe indicated along the main diagonal, $s_L= s_R$; away from this line
the finite polarization gradient $\Delta s=s_L - s_R$ mixes $\Ket{T_0}$ with the singlet states [see Fig.\ref{fig:TT} and Eq.(\ref{H3x3}) below] and gives rise to enhanced current, which is then only weakly sensitive to DNP. 

Intermittency originates naturally as follows.
Due to asymmetry of the dots, 
DNP initially moves from the unpolarized state into the region $s_L \neq s_R$ where current is insensitive to polarization (pink curve in Fig.\ref{fig1}b).  
This corresponds to the quiet build-up period in the current trace (see Fig.\ref{fig:exp_traces}). 
After approaching the nearly symmetric fixed point, DNP continues to fluctuate locally. 
Here, relatively small fluctuations of polarization 
take the system in and out of the low-current stripe $s_L \approx s_R$, resulting in large amplitude fluctuations of current and its apparent `switching' between high and low values. 

This behavior arises 
whenever a DNP fixed point resides near the sensitive region $s_L\approx s_R$, irrespective of the details of the dynamics near the fixed point.
Using a realistic model of the stochastic dynamics of electron transport and nuclear polarization, 
we have generated current traces exhibiting both the slow quiet build-up and long-time intermittent fluctuations. 
The relationship between the intermittent behavior of current and fluctuating DNP is illustrated in Fig.\ref{fig1}c, where corresponding regions of low current and $s_L \approx s_R$ are marked in gray. 
When parameters such as magnetic field or detuning are changed such that the DNP fixed point moves away from 
$s_L\approx s_R$, intermittent fluctuations of current are abruptly suppressed (see red line in Fig.\ref{fig:exp_traces}b); 
this picture 
is consistent with  experiment (see Fig.\ref{fig:exp_traces}a and Ref.~\onlinecite{Koppens}).




\section{Mathematical Formulation}
\label{sec:Model}
\begin{figure}[t]
\includegraphics[width=3.4in]{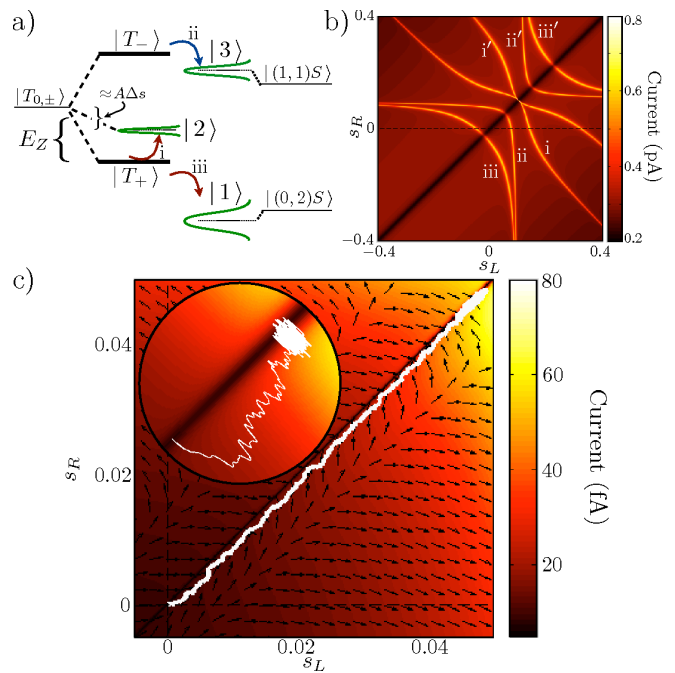}
 \caption[]{a) Energy level diagram showing hybridization of $S^z_{\rm Tot} = 0$ levels, see Eq.(\ref{H3x3}), and relevant transitions for spin-blockaded transport.  
b) Current as a function of nuclear polarizations $s_L$ and $s_R$ on the left and right dots.
Current is suppressed for $s_L \approx s_R$ 
due to the vanishing coupling between $\Ket{T_0}$ and $\Ket{(1,1)_S}$. 
Small roman numerals indicate bands of high current due to resonance between blocked and unblocked levels, shown in panel a).  Primed labels indicate analogous configurations with $\Ket{T_\pm}$ reversed.
c) Simulated trajectory of polarization arcing through the smooth region of the current diagram with steady-state near the sensitive region $s_L \approx s_R$, similar to the schematic in Fig.\ref{fig1}. 
Arrows indicate the direction of the flow (\ref{FlowEqn}). 
Inset: distorted zoom of a narrow strip containing the diagonal, magnified in the transverse direction to display the arc-shaped trajectory and steady state fluctuations.
 }
\label{fig:TT}
\end{figure}

We now turn to a detailed description of the model used to generate 
Figs.\ref{fig1} and \ref{fig:exp_traces}.
The relevant energy levels 
are depicted in Fig.\ref{fig:TT}. 
The states $(1, 1)_S$ and $(0,2)_S$, coupled by spin-conserving interdot tunneling with amplitude $t$, 
exhibit an avoided level crossing as a function of detuning $\Delta$.
A uniform magnetic field $B$ splits the $(1, 1)$ triplet into the states $\Ket{T_0}$, $\Ket{T_\pm}$ with total $z$-projection of electron spin $m = 0, \pm 1$, respectively. 

The hyperfine interaction gives rise to the Overhauser shift of the Zeeman energy, which is different on the left and right dots: $H^{Z}_{L(R)} = (g \mu_B B + A s_{L(R)}) S^z$.
The `polarization gradient' $\Delta s = s_L - s_R$ couples the states $\Ket{T_0}$ and $\Ket{(1, 1)_S}$. 
Including this coupling, we find the energy levels $\epsilon_n$ and eigenstates $\Ket{n}$ within the $S^z_{\rm Tot}= 0$ subspace (spanned by the states $\Ket{T_0}$, $\Ket{(1,1)_S}$, $\Ket{(0,2)_S}$) by diagonalizing the Hamiltonian 
\begin{eqnarray}
\label{H3x3} 
\hat{H}_{3\times 3} =
  \left(\begin{array}{ccc}
    0 & A \Delta s/2 & 0 \\
    A \Delta s/2 & 0 & t \\
    0 & t &  -\Delta - i\hbar\gamma/2
    \end{array}\right),
%
%
\end{eqnarray} 
where the energy of the unhybridized $(0, 2)$ singlet state, $\epsilon_{(0,2)_S} = -\Delta - i\hbar \gamma/2$, includes an imaginary part that accounts for its decay due to coupling to the continuum of states in the drain lead.
Due to a large applied bias, we assume that the $(1,1)$ state cannot decay directly to the continuum.
After diagonalization, each of the states $\Ket{n}$ obtains a nonzero $\Ket{(0,2)_S}$ component and decays with a rate $\gamma_n = -2{\rm Im}[\epsilon_n]/\hbar$, see Fig.\ref{fig:TT}a.

Current through the system results from electron transmission through any of 
the 5 states $\{\Ket{n}, \Ket{T_\pm}\}$ that may be populated when the second electron is injected from the lead.
The total current 
\begin{eqnarray}
\label{eq:I}  I(s_L, s_R; t, \Delta, B_0, \ldots) = \left(\sum_n p_n \tau_n + p_+ \tau_+ + p_- \tau_- \right)^{-1}
\end{eqnarray}
is determined by the inverse of the average of the lifetimes $\{\tau_{n,\pm}\}$ of these states, weighted by the probabilities $p_{n,\pm}$ to load each of the states.

Cotunneling or spin-exchange with the leads with rate $W_{\rm cot}$ adds an additional decay channel, leading to the inverse lifetimes $\tau_n^{-1} = \gamma_n + W_{\rm cot}$, see Refs.\onlinecite{RudnerDNP, Saito,Vorontsov,Qassemi}.
The inverse lifetimes of $\Ket{T_\pm}$ are determined by the rates $W_{\pm}$ of resonant hyperfine flip-flop transitions to each of the states $\{\Ket{n}\}$, Eqs. (\ref{WL}) and (\ref{WR}) below, and of cotunneling: $\tau_\pm^{-1} = W_{\pm} + W_{\rm cot}$.
We neglect spin-orbit coupling, which is weak in GaAs, but do not expect any qualitative changes if it is included. 

Figure \ref{fig:TT}b shows
current as a function of 
polarizations $s_L$ and $s_R$ on the two dots for a fixed set of external conditions. 
In the dark stripe along the main diagonal, $s_L = s_R$, current is set by the cotunneling rate $W_{\rm cot}$; away from this line 
the finite polarization gradient $\Delta s$ mixes $\Ket{T_0}$ with the singlet states and gives rise to the large red plateau of enhanced current that spans most of the figure.
The width of the dark stripe is set by the Overhauser field difference required to mix the $\Ket{T_0}$ state and the $\Ket{(1,1)_S}$-like hybridized singlet state, and depends on tunnel coupling and detuning, as well as the cotunneling rate $W_{\rm cot}$, which controls the saturation of current.
Sharp yellow bands of further enhanced current appear 
where the Overhauser shift brings either $\Ket{T_+}$ or $\Ket{T_-}$ into resonance with one of the states with $S^z_{\rm Tot} = 0$. 
With the help of Fig.\ref{fig:TT}a, each line can be identified with a specific resonant transition. 

As seen in Fig.\ref{fig:TT}b, 
the $(s_L, s_R)$ plane includes vast regions in which current is essentially independent of the value of polarization.
In 
these regions, the mean value of polarization cannot be inferred and fluctuations about the mean do not induce fluctuations of current.
In other regions, in particular near the line $s_L = s_R$, 
the derivatives $\partial I/\partial s_{L,R}$ are large; here current is very sensitive to small fluctuations of polarization which can result in large-amplitude intermittent fluctuations of current, as discussed above.

As discussed in Ref.\onlinecite{RudnerDNP}, DNP arises during spin-blockaded transport as the result of competition between hyperfine decay of $\Ket{T_\pm}$, which changes the z-component of nuclear spin, and non-spin flip decay channels.
Each time an electron in the state $\Ket{T_{+(-)}}$ decays by hyperfine spin flip, a nuclear spin is flipped from down to up (up to down).
By 
adding the transition rates from $\Ket{T_\pm}$ to all three final states in the decaying $S^z_{\rm Tot} = 0$ subspace, obtained using Fermi's Golden Rule, 
we find the ``bare'' transition rates $W_{\pm,L(R)}$ for hyperfine decay of $\Ket{T_\pm}$ assisted by a nuclear spin flip in the left (right) dot:
\begin{eqnarray}
    \label{WL} 
&& W_{\pm,L} = \!\sum_n  \!\frac{|a_n \mp b_n|^2}{16 N_L}\, \frac{A^2(1 \mp s_L)\,\gamma_n}{(\epsilon_\pm - {\rm Re}\,\epsilon_n)^2 + (\hbar \gamma_n/2)^2}
,\ 
\\
    \label{WR} 
&& W_{\pm,R} =   \! \sum_n \!\frac{|a_n \pm b_n|^2}{16 N_R} \, \frac{ A^2(1 \mp s_R)\,\gamma_n}{(\epsilon_\pm - {\rm Re}\,\epsilon_n)^2 + (\hbar \gamma_n/2)^2}
,
\end{eqnarray}
with the hybridized states
$\Ket{n}  = a_n\,\Ket{T_0} + b_n\,\Ket{(1, 1)_S} + c_n\,\Ket{(0, 2)_S}$
obtained from (\ref{H3x3}), and $\epsilon_\pm = g\mu_B B \pm A(s_L+s_R)/2$.
The states $\Ket{n}$ are characterized by a nonuniform electron spin density on the two dots, which introduces an asymmetry in the rates $W_{\pm,L(R)}$ to flip nuclear spins on the left and right dots via the dependence on 
$a_n$, $b_n$. 


The net spin flip rate 
in the left (right) dot is proportional to the total current $I$, Eq.(\ref{eq:I}), to the probability $p_\pm$ of loading $\Ket{T_\pm}$, and to the probability $W_{\pm,L(R)}/(W_{\pm,L}+W_{\pm,R} + W_{\rm cot})$ that this state, when loaded, decays by hyperfine spin flip in the left (right) dot, 
giving 
\begin{eqnarray}
\label{SFRates}  \Gamma_{\pm,L(R)}\ =\ p_{\pm}(I/e) \, \frac{W_{\pm,L(R)}}{W_{\pm,L}+W_{\pm,R} + W_{\rm cot}}.
\end{eqnarray}
Our separate treatment of $\Ket{T_\pm}$ and $\Ket{T_0}$ is valid in nonzero field where the degeneracy of these states is lifted.
Near zero field one should employ a $5\times 5$ generalization of Eq.(\ref{H3x3}) as in Ref.\onlinecite{Nazarov}.

Finally, using Eq.(\ref{SFRates}), and including relaxation with rate $\Gamma_{\rm rel}$, we arrive at the equations of motion for $s_L$ and $s_R$:
\begin{eqnarray}
  \label{FlowEqn}
  \begin{array}{rcl}
    \dot{s}_L &=& 2(\Gamma_{+,L} - \Gamma_{-,L})/N_L\ -\ \Gamma_{\rm rel}\,s_L\\
    \dot{s}_R &=& 2(\Gamma_{+,R} - \Gamma_{-,R})/N_R\ -\ \Gamma_{\rm rel}\,s_R,
  \end{array}
\end{eqnarray}
where 
$s_{L(R)}$ are defined as $(N_+-N_-)/N$ in each dot.

\section{Simulations and Results}
\label{sec:Results}
Equation (\ref{FlowEqn}) defines a flow, illustrated by the arrows in Fig.\ref{fig:TT}c, that describes the smooth dynamics of mean polarization.
However, polarization is actually stochastically driven by a train of electrons passing through the system, and executes a 
directed random walk around the flow (\ref{FlowEqn}).
 Fluctuations 
arise from shot noise in the number of entering up and down spins,
from the random competition between hyperfine and cotunelling decay channels, and from nuclear spin diffusion/relaxation.
We simulate this random walk by stochastically loading electrons into each of the 5 transport channels and then generating a corresponding sequence of randomly distributed decay times and numbers of nuclear spin flips, with mean values given by Eqs.(\ref{H3x3}), (\ref{WL}), and (\ref{WR}).

From the simulation we obtain current and polarization trajectories as shown in Figs.\ref{fig1}c and \ref{fig:exp_traces}b, see appendix for parameters.
For dots of unequal sizes, $N_L \neq N_R$, the flow (\ref{FlowEqn}) 
is asymmetric with respect to $s_L$ and $s_R$.
In particular, the rates (\ref{WL}) and (\ref{WR}) favor spin flips in the smaller of the two dots due to the increased hyperfine coupling per nuclear spin. 
As a result, the system can follow an arc-shaped trajectory like that shown in Figures \ref{fig1}b and \ref{fig:TT}c, in which polarization passes through the insensitive region $s_L \neq s_R$ during its build up, eventually returning to a steady state $s_L\approx s_R$ where polarization fluctuations result in large fluctuations of current.

Slow fluctuations with a power spectrum close to $1/f^2$, see Fig.\ref{fig:exp_traces}c, 
are indicative of diffusion, 
which may be driven by nuclear dipole-dipole interactions, as discussed in Ref.\onlinecite{Reilly}, 
or by current as described above. 
Unlike other sources of steady-state spin fluctuations, 
the shot-noise mechanism is intrinsic to spin-blockade and its intensity can be controlled by current. 
Alternative mechanisms can thus be distinguished through the current-dependence of the underlying diffusion coefficient.
\begin{figure}
\includegraphics[width=3.4in]{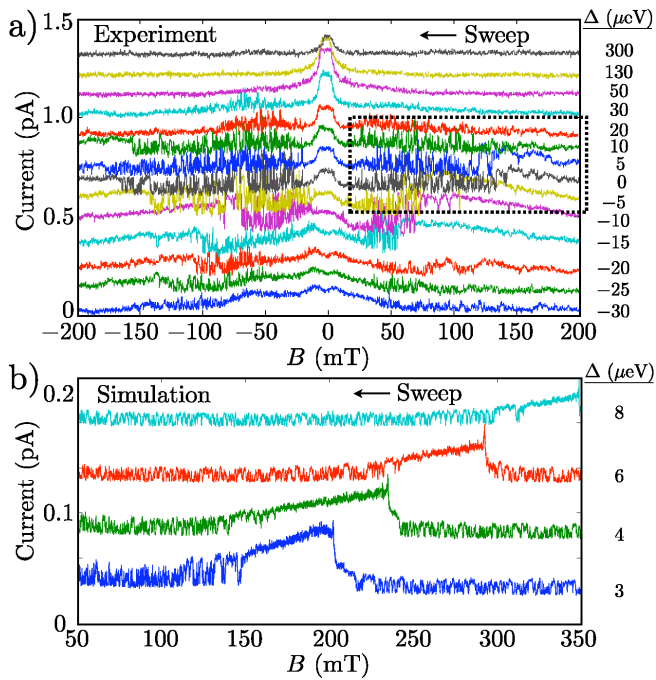}
 \caption[]{Current during magnetic field sweep $dB/dt < 0$.  a) Experimental field sweeps for several values of $\Delta$, see Eq.(\ref{H3x3}), in the regime described in Ref.\onlinecite{Koppens}, Fig. 3d ($\Delta = 0$ could only be approximately located experimentally). Curves are offset by 0.1 pA for clarity.
 b) Simulated field sweeps in the field regime indicated by the dotted box in panel a), showing bistable current similar to experiment.  
In both cases, the threshold for noisy to quiet current moves to higher field with increasing detuning.
To treat the case of $B\approx 0$, a $5\times 5$ formulation akin to that of Ref.\onlinecite{Nazarov} must be used. 
}
\label{fig:Sweeps}
\end{figure}

Experimentally, 
current was also measured during slow sweeps of magnetic field (see Fig.\ref{fig:Sweeps}a).
These data were obtained in the same regime as that of Fig.3D in Ref.\onlinecite{Koppens}, with a small change in tunnel coupling. 
At large $\Delta$, current displays a simple peak at small magnetic fields arising from mixing of the triplet levels with the $(1, 1)$ singlet by the random hyperfine field \cite{Nazarov}.
However, when detuning is comparable to the tunnel coupling (dotted box), 
the traces show diminished zero field peaks flanked by noisy regions exhibiting large fluctuations and stable regions of high current at higher fields. 
The boundary between noisy and stable high current systematically moves to higher field as detuning is increased, and is hysteretic in the sweep direction.

By including a time-dependent external field in the simulation, we produced the field sweep traces shown in Fig.\ref{fig:Sweeps}b.
The low-field boundary between noisy and quiet regions depends on detuning in a similar manner to that observed in the experiment, while on the high field side we find an additional current-step not observed within the range of available experimental data.
Based on the corresponding behavior of DNP in the simulations, we thus interpret the transition from noisy to stable current in the experiment as an indication that the polarization quasi-fixed point, $\dot{s}_{L,R} = 0$ in Eq.(\ref{FlowEqn}), moves away from the sensitive region $s_L \approx s_R$.

\section{Conclusions}
\label{sec:Conclusions}

The mechanism described above, based on spin dynamics driven by electron shot noise, provides a natural explanation for
the systematic observation of regions of stable and strongly fluctuating current.
We propose that regions of high, stable current, see e.g. $\Delta = -10\,\mu$eV, $B > 60$ mT in Fig.\ref{fig:Sweeps}, indicate that the system tends to an asymmetric fixed point with a sizable difference between the hyperfine fields in the two dots.  
As recently demonstrated, such states can be used to perform controlled manipulations of the double dot electron spin states\cite{Foletti}.

We gratefully acknowledge financial support from FOM, NWO and the ERC (L.V.), NSF grants DMR-090647 and PHY-0646094 (M.R.), and the Intelligence Advanced Research Projects Activity (IARPA), through the Army Research Office (L.V. and M.R.).


\appendix
\section{Simulation Parameters}
The main text describes the microscopic model used to generate the simulated current and polarization traces in Figures 1-4. 
The behavior exhibited by the model is sensitive to a number of parameters, many of which are not well characterized for the experimental system.
While on the one hand the existence of many parameters makes direct comparison to experiment more difficult, it can also be seen as a necessary consequence of the fact that such a wide variety of complex phenomena have been observed in this system. 
To this end, we have attempted to include the minimum number of ingredients necessary to produce the phenomena of slow quiet transients followed by steady state fluctuations.
We chose parameters with plausible values for realistic systems, see table \ref{ParamTable}, which led to simulated traces clearly demonstrating the phenomena of interest on field and time scales approximately comparable to those observed in experiments. 

In the table below, $N_L$ and $N_R$ are the numbers of nuclear spins in the left and right dots, $W_{\rm cot}$ is the cotunneling rate, $t$ is the tunnel coupling, $\Delta$ is the detuning, $\gamma$ is the decay rate of $(0, 2)_S$ [see Eq.(1), main text], $\Gamma_{\rm rel}$ is the phenomenological relaxation rate for nuclear polarization within the double dot, Eq.(6) of the main text, and $B$ is the magnetic field strength.

\begin{table*}[ht] \caption{Simulation Parameters} 
\centering      
\begin{tabular}{| c | c | c | c | c | c | c | c | c |}  
\hline\hline  
Figure & $N_L$ & $N_R$ & $W_{\rm cot}$ & $t$ & $\Delta$ & $\gamma$ & $\Gamma_{\rm rel}$ & $B$\\ [0.5ex] 
\hline                    
1c & $\ 4.8\times 10^6\ $ & $\ 5\times 10^6\ $ & $\ 2\times 10^3$ s$^{-1}\ $ & $\ $6 $\mu$eV$\ $&$\ 3\ \mu$eV$\ $&$\ 0.3\ \mu$eV$\ $&$\ 3.6\times 10^{-3}$ s$^{-1}\ $ &$\ 80$ mT$\ $ \\  
2b,c & $\ 0.75\times 10^6\ $ & $\ 1.25\times 10^6\ $ & $\ 1\times 10^4$ s$^{-1}\ $ & $\ $2 $\mu$eV$\ $&$\ 4\ \mu$eV$\ $&$\ 0.5\ \mu$eV$\ $&$\ 0.6\times 10^{-2}$ s$^{-1}\ $ &$\ 300$ mT$\ $ \\
3c & $\ 0.75\times 10^6\ $ & $\ 1.25\times 10^6\ $ & $\ 1\times 10^4$ s$^{-1}\ $ & $\ $2 $\mu$eV$\ $&$\ 4\ \mu$eV$\ $&$\ 0.5\ \mu$eV$\ $&$\ 0.6\times 10^{-2}$ s$^{-1}\ $ &$\ 400$ mT$\ $ \\
3b\footnote{Parameters for Fig.3b are chosen to most clearly display the sharp resonance features due to level crossings.} & $\ 1\times 10^6\ $ & $\ 2\times 10^6\ $ & $\ 1\times 10^6$ s$^{-1}\ $ & $\ $15 $\mu$eV$\ $&$\ 10\ \mu$eV$\ $&$\ 0.1\ \mu$eV$\ $&$\ {\rm N/A}\ $ &$\ 650$ mT$\ $ \\
4 & $\ 1.6\times 10^6\ $ & $\ 2\times 10^6\ $ & $\ 1\times 10^5$ s$^{-1}\ $ & $\ $1.5 $\mu$eV$\ $&$\ 3$--$8\ \mu$eV$\ $&$\ 0.1\ \mu$eV$\ $&$1.2$ s$^{-1}\ $&$\ 50$ mT--350 mT$\ $ \\ [1ex]
\hline     
\end{tabular} \label{ParamTable}  
\end{table*} 

\end{document}